# Estimación del Exponente de Hurst en Flujos de Tráfico Autosimilares

G. Millán [a,*]

[a] Departamento de Ingeniería Eléctrica, Universidad de Santiago de Chile, Avda. Ecuador 3519, Estación Central, Santiago, Chile.

**Resumen**

Este trabajo presenta, desarrolla y discute la existencia de un proceso con estructura de memoria de largo alcance, representativo de la independencia entre el grado de aleatoriedad del tráfico generado por las fuentes y del patrón de flujo exhibido por la red, en términos de una nueva variante algorítmica del estimador de máxima verosimilitud (MLE) de Whittle para el cálculo del exponente de Hurst ($H$) de las series temporales autosimilares de segundo orden estacionarias representativas de los flujos tanto de las fuentes individuales como de su agregación. Se discute además la problemática adicional introducida por el fenómeno de la localidad del exponente de Hurst, que se presenta cuando los flujos de tráfico se componen de diversos elementos con diferentes exponentes de Hurst. Se expone esta instancia con el ánimo de ser considerada como una forma nueva y alternativa para abordar el modelado y simulación del tráfico en las actuales redes de computadoras.

*Palabras Clave:*
Autosimilitud, Exponente de Hurst ($H$), Estimador de máxima verosimilitud (MLE), Estimador de Whittle, Memoria de largo alcance, Dependencia de largo alcance, Modelado de tráfico telemático.

## 1. Introducción

El posicionamiento y consolidación de Ethernet como estándar predominante en el campo de las redes de computadoras a niveles de coberturas tanto locales como extensas frente a tecnologías ya maduras y tradicionales como Frame Relay o ATM, son hechos que se explican a partir de sus principales características, a saber; compatibilidad e interoperatividad entre equipamientos Ethernet pertenecientes a distintas generaciones y por ende de velocidades distintas, independencia del direccionamiento IP, escalabilidad y capacidad de autoconfiguración, altas prestaciones y, sin lugar a dudas, por su consabida economía de escala.

Ethernet, inicialmente a 3 Mb/s, ha evolucionado de 10 Mb/s a 100 Gb/s en tres décadas y del uso de los simples puentes ideados para la interconexión de redes con idénticos protocolos a niveles físicos y de acceso al medio ha evolucionado hacia conmutadores N*100 Gb/s con capacidades de ruteo y gestión de aplicaciones.

Comprendidos en este continuo evolucionar que experimentan las redes Ethernet, se encuentran dos aspectos de especial interés y criticidad. El primero dice relación con el abandono del medio compartido half-dúplex original para dar paso a enlaces dedicados full-dúplex, mientras que el segundo trata su extensión: Ethernet ha evolucionado de distancias del rango LAN a coberturas WAN (Ibañez 2005). Y aun cuando ambos cambios, han sido graduales en el tiempo, son radicales desde el punto de vista topológico de Ethernet porque, en conjunto, han significado la desaparición del mecanismo de control de acceso al medio CSMA/CD e impuesto un drástico cambio en los de medios de transmisión tendiente al uso absoluto de fibras ópticas para brindar soporte a aplicaciones cada vez más demandantes de ingentes anchos de banda.

Las redes LAN, en general, y Ethernet en particular, nacieron siendo, en esencia, redes de medio compartido de alta capacidad, frente a tecnologías de redes WAN basadas en conmutación y con caudales de transmisión generalmente inferiores a los disponibles en las redes LAN. Sin embargo, la evolución de las tecnologías utilizadas en ambos entornos converge, hoy en día, en soluciones basadas en Ethernet y sus diversas especificaciones. Luego, las actuales redes LAN Ethernet son conmutadas, están compuestas casi en su totalidad por enlaces dedicados full-dúplex, incorporan la multiplexación basada en el estándar IEEE 802.1Q y soportan distancias de transmisión idénticas a las soportadas por los enlaces WAN convencionales. Este grado de evolución se atribute en gran medida al alto nivel de desarrollo alcanzado por los conmutadores Ethernet, el cual no tan solo se ha visto reflejado en el aumento de la transparencia y simplicidad operacionales, sino que además, ha incidido de manera directa en la incorporación de funcionalidades adicionales sobre la simple conmutación, lo cual desde el punto de vista del estándar, se ha traducido en la extensión del formato de trama incorporando etiquetados para VLAN y establecimiento de prioridades para clases de servicio (CoS), la supresión de la señal portadora de CSMA/CD para las actuales especificaciones; y para la necesaria convergencia entre generaciones, en la incorporación de ráfagas de paquetes para compensar las pérdidas de velocidad ocasionadas por los bits de extensión de la portadora y finalmente en el abandono del mecanismo para contención y resolución de colisiones determinado por el algoritmo de retroceso exponencial binario randómico (García, et al., 1997; Zacker, 2002).

---

* Autor en correspondencia.
  *Correo electrónico*: `ginno.millan@usach.cl` (G. Millán).

Por otra parte, la tendencia migratoria hacia redes Ethernet sin medio compartido queda confirmada con la incorporación de los estándares IEEE 802.1X, IEEE 802.1 D (RSTP, ex IEEE 802.1w) e IEEE 802.1Q (MSTP, ex IEEE 802.1s), los cuales requieren de enlaces dedicados full-dúplex para su operación. Al respecto cabe señalar que estos enlaces no tan solo son necesarios para obtener las máximas prestaciones de la red, sino que además posibilitan la seguridad a nivel de enlace lógico de datos (LLC), simplifican sus protocolos y procesos y permiten la operación de los mecanismos de convergencia rápida en capa dos.

Un ejemplo categórico de la discusión anterior lo constituye el estándar IEEE 802.3ae (Ethernet a 10 Gb/s), el cual no contempla en su especificación el uso de enlaces half-dúplex como ocurre en IEEE 802.3z (Ethernet a 1 Gb/s), donde su mantención obedece a estrictas razones de compatibilidad con las bases de equipamiento instaladas con anterioridad, teniendo como objetivo final, el servir como plataforma en procesos de migración tecnológica. Llegado este punto es necesario aclarar que IEEE 802.3z es la última de las especificaciones del proyecto 802.3 que brinda soporte a esta modalidad de comunicación entre dispositivos.

Resulta de gran atención e interés el hecho de que en las redes Ethernet tradicionales predominasen la difusión y la inundación como mecanismos básicos y válidos al momento de establecer la presencia o la ausencia de estaciones, y que hoy, por el contrario, se busque la mínima difusión de las tramas por las mismas dos razones por las que se evita en los entornos WAN: la degradación del rendimiento y el control exhaustivo de los flujos de tráfico.

Finalmente se argumenta que paralelamente a todo lo anterior las redes de mayores coberturas están incorporando tecnologías propias de entornos LAN debido a su robustez y buena relación precio/prestaciones estando ya bastante implantadas en entornos tanto de acceso como metropolitanos y de manera creciente en los entornos WAN propiamente tales (Halabi, 2003).

Todos los argumentos anteriores justifican el planteamiento de un nuevo enfoque para llevar a cabo los procesos de modelado de redes Ethernet, puesto que en términos de la evolución planteada, se infiere que el sucesor del estándar IEEE 802.3u (Fast Ethernet), es el estándar IEEE 802.3z (Gigabit Ethernet), que cederá su lugar en los entornos de acceso y WAN a los estándares IEEE 802.3ae (Ethernet 10 Gb/s) e IEEE 802.3ba (Ethernet 40 Gb/s y 100 Gb/s), respectivamente, y los efectos que estas migraciones tecnológicas involucren deben ser adecuadamente dimensionados, evaluados y clasificados en función de su impacto sobre el rendimiento de las bases de equipamiento instaladas. En este sentido, el rendimiento, entendido como la razón entre la cantidad de información útil y la cantidad total de bits que transporta la red, se considera como una forma activa de medir las prestaciones, puesto que es uno de los aspectos de mayor interés en el análisis global de los sistemas de comunicaciones debido a las repercusiones que posee sobre los usuarios finales en términos de su percepción del sistema. Luego, resulta de vital importancia modelar adecuadamente el tráfico de entrada a las redes por cuanto de él depende el comportamiento que exhiba el parámetro de rendimiento en cada caso particular, convirtiéndose en un factor clave para su estimación. Caracterizar adecuadamente el tráfico de entrada a una red involucra describir estadísticamente tanto su naturaleza como su evolución temporal a partir de las series de datos que lo representan.

Aceptando la naturaleza autosimilar de los flujos de tráfico en Ethernet; comprobada por la presencia permanente de ráfagas de paquetes con independencia de la escala temporal de observación, y su comportamiento con dependencia de largo alcance (LRD); el cual se caracteriza por una función de autocorrelación cuyo valor se mantiene constante para los distintos niveles de agregación, se hace necesario estudiar los valores que, para las series temporales representativas de los flujos, presenta el exponente de Hurst ($H$), el cual, por definición, expresa la velocidad de decrecimiento de la función de autocorrelación; en otras palabras mide la rugosidad en el curso de las muestras y, por lo tanto, captura la autosimilitud de las series. Existen diversos métodos para calcular el exponente de Hurst, siendo el del estimador de máxima verosimilitud (MLE) de Whittle (Whittle, 1953) el más ampliamente utilizado debido a su mayor rigor estadístico comparado con los métodos gráficos de varianza agregada (Var) o de rango re-escalado (R/S) y, porque, además, proporciona la varianza de la estimación, posibilitando la obtención de intervalos de confianza. Esto último, debido a que el MLE de Whittle es asintóticamente normal.

Cabe señalar que en el estimador de Whittle se basan una serie de métodos para estimar parámetros mediante MLE, los cuales en su totalidad, incluido el de Whittle, requieren conocer, en mayor o menor grado, la forma paramétrica del espectro de frecuencias del proceso estocástico objetivo. Los estimadores de Whittle discreto (Graf, 1983), agregado y local (Taqqu and Teverovsky, 1997) son las realizaciones más reconocidas y utilizadas. Lamentablemente, el común denominador, para todos los métodos basados en MLE, es su alto costo computacional de implementación. Y es más, aun cuando los últimos tres métodos mencionados logran minimizar de buena manera, analíticamente, las funciones objetivo, el problema de los altos tiempos de cómputo persiste, y acrecienta para el caso de las series relacionadas con los tráficos en las actuales redes de computadoras de alta velocidad producto de sus extensiones y a la elevada complejidad derivada de sus fluctuaciones localizadas.

Luego, considerando los argumentos anteriores, en este trabajo se presenta una nueva variante del MLE de Whittle para estimar el valor del exponente $H$ de series autosimilares de segundo orden estacionarias, representativas de los flujos de tráfico presentes en las actuales redes de computadoras de alta velocidad, a través de un algoritmo que minimiza la cantidad de puntos a evaluar y que aprovecha la convexidad de la función objetivo en el intervalo de interés [0.5, 1[; entregando una relación aceptable de compromiso entre el costo computacional (el mayor inconveniente presente en la obtención del MLE de Whittle independientemente del método elegido) y la calidad de las estimaciones.

La propuesta consiste en modificar el MLE local de Whittle, o estimador gaussiano semiparamétrico del parámetro de memoria en procesos estándar a largo plazo original de Robinson (1995), con el ánimo de conseguir las ventajas de la técnica original; que muestra sus principales atributos como alternativa a la técnica de regresión del logaritmo del periodograma de Geweke and Porter-Hudak (1983). Particularmente, se espera que, con supuestos aun menos restrictivos, se muestre una forma de ganancia asintótica al valor final de $H$. El procedimiento general consiste, primeramente en analizar el comportamiento asintótico del estimador gaussiano semiparamétrico original del parámetro de memoria en procesos con memoria cíclica o estacional, permitiendo, así, divergencias (o ceros espectrales asimétricos), para luego, por intermedio de la modificación del algoritmo original, obtener la consistencia y la normalidad asintótica necesarias para lograr caracterizar de forma adecuada los flujos de tráfico objeto de estudio en términos de sus series temporales subyacentes.

## 2. Planteamiento del Problema

El análisis de colas ha resultado ser de enorme utilidad para el diseño de redes y el análisis de sistemas, a efectos de llevar a cabo planificaciones de capacidades y predicción de rendimientos. Sin embargo, muchos casos del mundo real, colocan en evidencia que los resultados predichos a partir de un análisis de colas difieren de manera significativa del rendimiento observado en la realidad. En este contexto, se recuerda que la validez de los análisis basados en teoría de colas dependen de la naturaleza de Poisson del tráfico de datos (Stallings, 2004) y que al tratarse de procesos de Poisson, la representación tanto de la duración de los arribos sucesivos como del tiempo entre ellos, está dada en función de variables aleatorias independientes exponencialmente distribuidas. Por este motivo, se trata de modelos con memoria nula y, por ende, de modelos en los cuales la probabilidad de ocurrencia de una llegada en un instante determinado no depende de las llegadas anteriores; propiedad que se incumple en las redes de conmutación de paquetes, no obstante se reconoce que el objetivo de ambos supuestos tan solo responde a obtener modelos de tráfico relativamente simples desde el punto de vista del tratamiento analítico de los mismos.

Desde el establecimiento de la teoría matemática que gobierna el comportamiento de las redes de conmutación de paquetes por parte de Kleinrock (Kleinrock, 1961), posteriormente ampliada y detallada en (Kleinrock, 1972), la existencia de las dependencias temporales en los comportamientos de los flujos de tráfico en las redes de comunicaciones se ha convertido en un área de candente investigación, contando a su haber innumerables publicaciones y descubrimientos, destacándose entre estos últimos uno de especial interés y criticidad para el planteamiento de esta investigación; el del impacto e influencia que las dependencias temporales poseen sobre las prestaciones de los sistemas de colas, por cuanto, a raíz de ello, es que se explican tanto la existencia como la coexistencia actuales de una amplia gama de modelos de tráfico de entrada que presentan estructuras con correlaciones más o menos complejas, aplicados a casos en los cuales los modelos de comunicaciones bajo estudio permiten mantener un manejo analítico adecuado. En todos los casos, estos modelos, fundamentalmente Markovianos, desprecian las correlaciones temporales a partir de determinada escala aun cuando pueda ser arbitrariamente aumentada a costa de complicar los modelos con parámetros adicionales.

Desde 1993, un creciente número de estudios publicados en la literatura han documentado que el patrón de flujos de tráfico se modela bien mediante procesos autosimilares en una amplia gama de situaciones pertenecientes al mundo real y de las redes, dando cuenta de (Leland *et al.*, 1993), luego aumentada y corregida en (Leland *et al.*, 1994), como obra fundacional y referente donde se demuestra, en lo fundamental, después de exhaustivas mediciones de tráfico, que el tráfico en Ethernet posee naturaleza autosimilar, o fractal, con independencia del momento y lugar de su medición, la cual se manifiesta a través de una acusada correlación de largo alcance, o dependencia de largo alcance (LRD). Una recopilación de investigaciones consideradas basales a continuación prosigue.

El comportamiento autosimilar LRD de los flujos de tráfico en las actuales redes de computadoras de alta velocidad es un hecho ampliamente reportado para topologías de diferentes niveles de coberturas (Leland *et al.*, 1994; Paxson and Floyd, 1995; Crovella and Bestavros, 1997; Arlitt and Jin, 2000); tecnologías y técnicas de transmisión (Ge *et al.*, 2000; Liang, 2002; Huang *et al.*, 2005; Fantacci and Tarchi, 2006; Huang *et al.*, 2007; Maier, 2008); protocolos de control y de señalización (Duffy *et al.*, 1994; Pruthi and Erramilli, 1995); aplicaciones de video, especialmente de tasa de bits variable, (Garrett and Willinger, 1994; Beran *et al.*, 1995; Yellanki, 1999, Narasimha and Rao, 2002; Narasimha and Rao, 2003), sistemas de colas en redes ATM (Norros, 1994; Likhanov *et al.*, 1995; Tsybakov and Georganas, 1997; Jin *et al.*, 2002); y en comunicaciones inalámbricas (Yu, 2005; Ridoux *et al.*, 2006, Radev and Lokshina, 2009; Yeryomin and Petersons, 2011).

Así mismo, el problema del modelado del tráfico de entrada a las actuales redes de comunicaciones es un tema con una amplia cobertura bibliográfica, que ha resultado en una amplia variedad de modelos, entre los que se encuentran, sin constituir un listado exhaustivo; procesos de Poisson con conmutación generalizada (Tran-Gia, 1983), procesos de Markov modulados por Poisson (Heffes and Lucantoni, 1986), procesos de Poisson conmutados (Rossiter, 1988), procesos fractales puntuales (Ruy and Lowen, 1996), procesos con renovación fractal alternante y renovación fractal alternante extendida (Willinger *et al.*, 1997; Yang, 2001), procesos basados en la iteración de mapas caóticos intermitentes (Erramilli *et al.*, 1995; Ding and Yang, 1995; Mondragón *et al.*, 2000; Arrowsmith *et al.*, 2004), y los tradicionales procesos de ruido y movimiento gaussiano fraccionales (Millán and Lefranc, 2013, Millán, 2013xson, 1997).

Lo interesante de estos y otros muchos estudios es que colocan de manifiesto el impacto que la autosimilitud LRD inherente a los más diversos tipos de tráfico, posee sobre las prestaciones de las redes de comunicaciones, frente a modelos que ya sea por atender la simplicidad analítica o porque sus estructuras de correlaciones temporales son mucho más complejas, no la contemplan. En este respecto, todos los modelos englobados dentro de las últimas dos frases, se conocen como modelos tradicionales o con memoria de corto alcance (SRD); cuyo principal inconveniente radica en que requieren de un muy elevado número de parámetros de modelado para caracterizar de forma adecuada las fuertes correlaciones que existen entre los diferentes tipos de tráfico presentes en las redes. En este punto se hace hincapié en que al aumentar la cantidad de parámetros de modelado, aumenta la complejidad de tratamiento analítico y no necesariamente en proporción lineal. Ello, además, sin considerar la dificultad añadida de proveer una interpretación física plausible a estos parámetros adicionales y lograr estimarlos adecuadamente a partir de datos empíricos.

En términos generales, la autosimilitud puede entenderse como el fenómeno evolutivo por la cual una determinada propiedad de un objeto se preserva ante sucesivos escalamientos temporales y/o espaciales; y un objeto autosimilar (fractal), es aquel que registra como singularidad el que la mínima expresión escalada de alguna de sus partes magnificada, se asemeja, en formas, al objeto en su totalidad. Semejanza dimensionada, ciertamente, en algún sentido estadístico adecuado. Una muy interesante discusión sobre el uso de estos términos se encuentra en Taqqu *et al.* (1997). Luego, por medio de un proceso de iteración es posible obtener la forma más sencilla de autosimilitud; si este proceso puede repetirse de forma indefinida para cualquier nuevo segmento, con independencia de su ubicación temporal y/o espacial, entonces cualquier porción del objeto es candidata a magnificarse en escala para representar con exactitud a cualquier otra parte del objeto por mayor que esta sea. Esta propiedad se conoce como autosimilitud exacta (Millan and Lefranc, 2009). Resulta evidente que no es posible observar este comportamiento en series temporales representativas de flujos de tráfico en redes de computadoras, pero si el tráfico observado se considera como trazas de muestras de un proceso estocástico y se restringe la similitud a estadísticos específicos ajustados a escala, se tendrá: autosimilitud exacta en objetos matemáticos abstractos y, aproximada en las realizaciones específicas que se consideren.

Declarado lo anterior, se espera que las series temporales bajo estudio muestren un comportamiento autosimilar estocástico y en este escenario, para determinar la autosimilitud, usar estadísticos de segundo orden que capturen la variabilidad de las muestras. Al respecto, por ejemplo, la invariancia ante los cambios de escalas puede definirse en términos de la función de autocorrelación si se argumenta que esta presenta un decrecimiento polinómico, en vez de exponencial, dado que ello es una clara manifestación de LRD, equivalente a autosimilitud. Con respecto a esto último se precisa que autosimilitud y LRD son conceptos diferentes. Sin embargo, en el caso de la autosimilitud de segundo orden, la autosimilitud implica LRD y viceversa (Yang and Petropulu, 2001).

Por otra parte, frente al problema de estimar $H$, las propuestas en la literatura pueden clasificarse en dos grandes grupos:

1. Métodos gráficos. Con ellos se calcula algún estadístico $T(x)$ que se comporta asintóticamente para un conjunto de valores de $x$, y se basan en obtener la recta que mejor se ajuste para dicho conjunto sobre la gráfica de $\log[T(x)]$ frente a $\log(x)$. El valor de $H$ se extrae de la pendiente de dicha recta.
2. Métodos que emplean estimadores de máxima verosimilitud (MLE). Con ellos se consigue minimizar las diferencias entre el periodograma y el espectro teórico de la serie en cuestión. El valor de $H$ se extrae directamente del algoritmo.

Los métodos pertenecientes al primer grupo son relativamente simples y fáciles algorítmicamente de implementar, presentando como principal inconveniente el que en un primer término debe estimarse un comportamiento asintótico a partir de una cantidad reducida de muestras, lo cual redunda en que la estimación de $H$ dependa directa y considerablemente de la correcta elección del conjunto de valores de $x$. Luego, por este preciso motivo, resultan fundamentales las representaciones gráficas para comprobar que el conjunto de valores elegidos para $x$ se corresponda con la zona de comportamiento lineal de la gráfica de $\log[T(x)]$ versus $\log(x)$, y que la recta obtenida es un buen ajuste para la representación. Estos métodos solo permiten obtener una estimación puntual de $H$, ya que la obtención de intervalos de confianza involucra un coste computacional adicional en términos de tiempos de proceso.

Por otra parte, los métodos basados en MLE, aun cuando más complejos y con un coste computacional mayor, son más flexibles y eficientes desde el punto de vista estadístico inferencial, puesto que permiten obtener intervalos de confianza en el cálculo de $H$.

En esta investigación se considera el uso de métodos basados en MLE para resolver los intervalos de confianza que conllevan a la obtención de un primer valor aproximado para $H$, el cual, luego de aplicado a la algorítmica de la metodología propuesta, resultará en un valor confiable. Métodos pertenecientes al primer grupo se consideran como pruebas de contraste, en específico el estadístico de rango re-escalado (R/S) y el de varianza-tiempo (V-T) (Zhang *et al.*, 1997, Pacheco, 2006; Millán *et al.*, 2016).

Como anteriormente se estableció, el cálculo de $H$ empleando métodos basados en MLE es computacionalmente costoso y ello se debe a su objetivo fundamental: minimizar las diferencias entre el periodograma de la serie y el modelo paramétrico supuesto para su densidad espectral teórica. Al respecto, un camino alternativo consiste en utilizar funciones de verosimilitud gaussianas (MLE gaussianos). Aun así el costo computacional sigue siendo elevado y por ello, en la práctica, se emplean aproximaciones basadas en dichos MLE gaussianos siendo la más ampliamente aceptada por su rigurosidad, la de Whittle. Por este motivo el método propuesto se basa en la modificación de una de sus variantes.

## 3. El Estimador de Whittle y sus Variantes

Sea $f(\lambda,\boldsymbol{\theta})$ la forma paramétrica de la densidad espectral de un proceso covariante estacionario $x_t$, donde: $\boldsymbol{\theta} = (\theta_1, \theta_2,\ldots, \theta_m)$ es el vector de parámetros que se desea estimar, $I(\lambda)$ el periodograma de $x_t$ y $\omega(\lambda)$ su transformada discreta de Fourier, tales que

$$I(\lambda) = |\omega(\lambda)|^2, \quad \omega(\lambda) = \frac{1}{(2\pi n)^{\frac{1}{2}}} \sum_{t=1}^{n} x_t e^{it\lambda}. \quad (1)$$

Luego, el MLE aproximado de Whittle es el vector

$$\hat{\boldsymbol{\theta}} = (\hat{\theta}_1, \hat{\theta}_2,\ldots, \hat{\theta}_m), \quad (2)$$

que minimiza la función

$$Q(\theta) = \frac{1}{2\pi} \left[ \int_{-\pi}^{\pi} \frac{I(\lambda)}{f(\lambda,\boldsymbol{\theta})} d\lambda + \int_{-\pi}^{\pi} \log f(\lambda,\boldsymbol{\theta}) d\lambda \right]. \quad (3)$$

En la práctica el estimador de Whittle se realiza escogiendo un parámetro de escala adecuado, $\theta_1$, tal que

$$f(\lambda,\boldsymbol{\theta}) = \theta_1 f(\lambda,\boldsymbol{\theta}^*) = \theta_1 f^*(\lambda,\boldsymbol{\eta}), \quad (4)$$

de forma que anule al segundo sumando de (3), es decir

$$\int_{-\pi}^{\pi} \log f(\lambda,\boldsymbol{\theta}) d\lambda = \int_{-\pi}^{\pi} \log[\theta_1 f^*(\lambda,\boldsymbol{\eta})] d\lambda = 0, \quad (5)$$

donde $\boldsymbol{\eta} = (\theta_1, \theta_2,\ldots, \theta_m)$ y $\boldsymbol{\theta}^* = (1, \boldsymbol{\eta})$.

En Beran (1994) se demuestra que el parámetro de escala debe ser $\theta_1 = \sigma^2 / 2\pi$, donde $\sigma^2$ es el error mínimo de predicción.

Por otra parte, en Graf (1983) se propone una versión discreta del estimador de Whittle que aproxima a (3) mediante una suma de Riemann en las frecuencias $\lambda_k = 2\pi n^{-1} k$, con $k = 1, 2,\ldots, n^*$ ($n^*$ es la parte entera de $(n - 1)/2$). Luego, la función a minimizar es

$$\tilde{Q}(\theta_1, H) = \frac{4\pi}{n} \left[ \sum_{k=1}^{n^*} \frac{I(\lambda_k)}{f(\lambda_k, \theta_1, H)} + \sum_{k=1}^{n^*} \log[f(\lambda_k, \theta_1, H)] \right]. \quad (6)$$

Eligiendo adecuadamente el parámetro de escala, el parámetro estimado para $H$, $\hat{H}$, es el valor que minimiza a

$$\tilde{Q}^*(H) = \tilde{Q}(1,H) = \sum_{k=1}^{n^*} \frac{I(\lambda_k)}{f(\lambda_k,1,H)} = \sum_{k=1}^{n^*} \frac{I(\lambda_k)}{f^*(\lambda_k,H)}, \quad (7)$$

donde

$$f^*(\lambda,H) = \frac{1}{\theta_1} f(\lambda,\theta_1,H) = \frac{2\pi}{\sigma_\varepsilon^2} f(\lambda,\theta_1,H) \quad (8)$$

y

$$\hat{\sigma}_\varepsilon^2 = 2\pi\hat{\theta}_1 = 4\pi n^{-1} \hat{Q}^*(\hat{H}). \quad (9)$$

Se advierte que las principales desventajas que el estimador de Whittle presenta en su forma convencional son:

1. Conocer la forma paramétrica de la densidad espectral.
2. Un mayor tiempo de cálculo que los métodos gráficos.

Desde la perspectiva de la aplicación del estimador de Whittle a procesos donde se hace imposible asegurar algo con respecto a su densidad espectral, resulta de gran interés el teorema del límite central para procesos autosimilares de Samorodnitsky and Taqqu (1994), el cual, dada su relevancia, se reproduce como sigue. Sea $X_j$ una serie gaussiana estacionaria de media cero. Según el valor de $H$ su función de autocovarianza, $\gamma(k)$, satisface:

- Si $1/2 < H < 1$, entonces

$$\gamma(k) \sim Ck^{2H-2}, \text{ cuando } k \to \infty \text{ con } C > 0. \quad (10)$$

- Si $H = 1/2$, entonces

$$\sum_{k=1}^{\infty} |\gamma(k)| < 0, \quad \sum_{k=-\infty}^{\infty} \gamma(k) = C > 0. \quad (11)$$

- Si $0 < H < \frac{1}{2}$, entonces

$$\gamma(k) \sim Ck^{2H-2}, \text{ cuando } k \to \infty, C > 0 \text{ y } \sum_{k=-\infty}^{\infty} \gamma(k) = 0. \quad (12)$$

El interés antes mencionado sobre este teorema radica en que resulta en una buena aproximación para series no Gaussianas; de hecho, el teorema permite suponer que para una serie temporal de tamaño $n$, con función de autocorrelación con caída hiperbólica con LRD; si $m$ y $n/m$ son suficientemente grandes y la varianza es finita, el proceso de ruido Gaussiano fraccional (FGN) constituye una buena aproximación para las secuencias agregadas de dicha serie (Taqqu and Teverovsky, 1998). Esta modificación conocida como estimador de Whittle agregado ofrece una manera de hacer más robusto y menos sesgado al estimador de Whittle cuando no se dispone de información acerca de la forma paramétrica exacta de la densidad espectral. Sin embargo, su principal inconveniente es que solo puede ser utilizado si la serie es suficientemente larga. Para ello se agregan los datos a través del proceso de agregación

$$X_k^{(m)} = \frac{1}{m} \sum_{i=km}^{(k+1)m-1} X_i, \quad 0 \leq k \leq \lfloor n/m \rfloor \quad (13)$$

con $\lfloor \cdot \rfloor$ parte entera, para obtener una serie más corta y sobre ella aplicar el estimador aproximado de Whittle considerando como modelo paramétrico de su densidad espectral el de FGN.

Quedan en evidencia dos situaciones de atención, a saber:

1. A pesar que el uso de (13) reduce considerablemente el costo computacional, presenta el inconveniente de que aumenta la varianza del estimador y por lo tanto disminuye el grado de representatividad del patrón.
2. Es imposible conocer a priori el valor apropiado de $m$, aun cuando en Leland *et al.* (1994) se propone una buena idea al representar $H$ para distintos valores de $m$ y luego buscar una región donde la gráfica sea aproximadamente plana.

En Robinson (1995) se propone el estimador de Whittle local o estimador gaussiano semiparamétrico. Esta técnica se denomina semiparamétrica, según Palma (2007), debido a que:

1. No requiere la especificación de un modelo paramétrico para los datos,
2. Sólo se basa en la especificación de la densidad espectral de las series de tiempo.

Asumiendo un proceso estacionario $x_t$ cuya densidad espectral satisface

$$f(\lambda) \sim G\lambda^{1-2H}, \text{ cuando } \lambda \to 0^+ \quad (14)$$

con $G \in ]0, \infty[$ y $H \in ]0, 1[$. Se verifica que:

- Si $1/2 < H < 1$, entonces $f(\lambda) \to \infty$.
- Si $H = 1/2$, entonces $f(\lambda) \to C \in ]0, \infty[$ a frecuencia cero.
- Si $0 < H < 1/2$, entonces $f(\lambda) \to 0$.

Considerando la función objetivo propuesta en Künsch (1987)

$$Q(G,H) = \frac{1}{m} \sum_{j=1}^{m} \left( \log G\lambda_j^{1-2H} + \frac{\lambda_j^{2H-1}}{G} I_j \right), \quad (15)$$

donde $I_j = I(\lambda_j)$ tal que, por inferencia directa de (1),

$$I(\lambda_j) = |\omega(\lambda_j)|^2, \quad \omega(\lambda_j) = \frac{1}{(2\pi n)^2} \sum_{t=1}^{n} x_t e^{it\lambda_j}, \quad j = 1,\ldots,m, \quad (16)$$

con $m$ entero tal que $m < n/2$, y las siguientes dos observaciones propuestas en Robinson (1995)

1. No es necesaria la corrección de una media desconocida de $x_t$ porque los estadísticos (1) se calculan solo en las frecuencias $\lambda_j = 2\pi j/n, j = 1,\ldots, m$, con $m$ entero menor que $n/2$, y
2. Puesto que la estimación (14) no está definida en una forma cerrada, conviene indicar por $G_0$ y $H_0$ los verdaderos valores de los parámetros, y por $G$ y $H$ a cualquier valor admisible.

Entonces, definido el intervalo de las estimaciones admisibles de $H_0$, $\Theta = [\Delta_1, \Delta_2]$, donde $\Delta_1$ y $\Delta_2$ son números elegidos de forma tal que $0 < \Delta_1 < \Delta_2 < 1$, es posible elegir $\Delta_1$ y $\Delta_2$ arbitrariamente cerca de 0 y 1, respectivamente, o bien se puede optar por reflejar los escasos conocimientos previos sobre $H_0$, por ejemplo, $\Delta_1 = 1/2$ si se está seguro que $f(\lambda) \not\to 0$ cuando $\lambda \to 0$. Luego, la estimación

$$(\hat{G}, \hat{H}) = \arg\min_{0 < G < \infty, H \in \Theta} Q(G,H) \quad (17)$$

claramente existe, y es posible escribir

$$\hat{H} = \arg\min_{H \in \Theta} R(H), \quad (18)$$

donde

$$R(H) = \log \hat{G}(H) - (2H-1) \frac{1}{m} \sum_{j=1}^{m} \log \lambda_j, \quad (19)$$

con

$$\hat{G}(H) = \frac{1}{m}\sum_{j=1}^{m}\lambda_j^{2H-1}I_j. \quad (20)$$

En Taqqu and Teverovsky (1997) se porta el mismo resultado para (19) a partir de considerar $R(H) = Q(\hat{G}, H)$.

Si $(\hat{G}, \hat{H})$, de (17), es el valor que minimiza a $Q(G, H)$, de (15), entonces bajo algunas condiciones de regularidad como

$$(m)^{-1} + (m/n) \to 0, \quad \text{cuando } n \to \infty, \quad (21)$$

Robinson (1995) establece el siguiente teorema (Palma, 2007): si $H_0$ es el valor verdadero del parámetro autosimilar, entonces el estimador $\hat{H}$ es consistente y

$$m^{1/2}(\hat{H} - H_0) \to N(0, 1/4), \quad \text{cuando } N \to \infty. \quad (22)$$

Este último resultado puede generalizarse como

$$m^{1/2}(\hat{H} - H_0) \to N(0, \sigma_H), \quad \text{cuando } N \to \infty; \; \sigma_H^2 = \frac{1}{4m}, \quad (23)$$

donde $\hat{H}$ es el valor del parámetro $H$ que minimiza a $R(H)$ de (19), $H_0$ es el valor verdadero del parámetro $H$ y $\hat{H} \to H_0$.

Se observa que, al igual que en el método de Whittle agregado, la elección del valor de $m$ resulta fundamental, apareciendo así el habitual compromiso entre el sesgo y la varianza. De hecho, en la medida que $m$ aumenta, $\hat{H}$ converge más rápidamente a $H_0$ pero, en cambio, la forma del espectro se aparta cada vez más de (14) y los efectos de la SRD serán cada vez mayores por lo que el sesgo aumentará. También, al igual que en los otros métodos tratados es conveniente representar $\hat{H}$ versus $m$ y encontrar la región plana de la gráfica. En Taqqu and Teverovsky (1997) se reporta que con un valor de $m = n/32$, utilizando series de 10000 muestras, se obtiene un compromiso aceptable entre sesgo y precisión.

## 4. Un Estimador Simple y Eficiente del Exponente de Hurst

Antes de plantear un nuevo algoritmo para el MLE de Whittle, conviene considerar los siguientes resultados relevantes:

1. En Taqqu and Teverovsky (1997) se analizan, entre otros, los métodos basados en el MLE de Whittle concluyendo que este último es la opción más adecuada cuando se elige el modelo correcto. En el caso de no conocerlo, los métodos de Whittle agregado y de Whittle local obtienen buenas aproximaciones y son computacionalmente más rápidos, sobre todo en el caso de estimador de Whittle local.

2. En Taqqu and Teverovsky (1998) se analizan, entre otros, los métodos tratados en la sección anterior y se observa, usando trazas F-ARIMA (1, $d$, 1) con diversas distribuciones para las innovaciones, que todos ellos son robustos con respecto a la distribución de las innovaciones. Además, con respecto a la presencia de SRD, los autores concluyen que los estimadores se ven fuertemente afectados por las componentes AR y MA, excepto el MLE de Whittle si el modelo usado es el correcto. En caso de no disponer de tal información, dicho estimador es el más sesgado y se recomienda el uso de los métodos de Whittle agrado o de Whittle local.

Estos resultados son interesantes puesto que como es posible observar de lo expuesto en la sección anterior, todos los métodos tratados requieren la minimización de una expresión ((7) o (19)), que para efectos de esta investigación es función exclusivamente de $H$. Luego, este último hecho da pie a suponer que la forma más obvia de realizar dicha minimización, consistiría en evaluar estas funciones para un cierto número de valores de $H$ equidistantes, $q$, que dependerá de la resolución deseada. Sin embargo, se advierte que un número moderadamente elevado de muestras provoca que el algoritmo se traduzca en un costo computacional muy elevado.

Para reducir este costo computacional se propone un algoritmo de minimización que reduce la cantidad de puntos a evaluar sobre la misma metodología, aprovechando que la función objetivo, (7) o (19), es convexa en todo el dominio [0.5, 1[, y que, además, su mínimo es único por este último hecho. Entonces, bajo el amparo de ambos hechos, un método de búsqueda por bisección, aplicado sobre la derivada de la función, permite que el número de puntos evaluados esté en entorno a $\log_2(q)$, lo cual redunda en un ahorro significativo de muestras manteniendo un compromiso aceptable entre el sesgo y la calidad de las estimaciones. Para producir estos efectos, se sabe que la derivada de una función, $f$, evaluada en un punto, $a$, puede aproximarse por un cociente de diferencias para un incremento suficientemente pequeño, $h$, mediante

$$f'(a) = \lim_{h \to 0}\frac{f(a+h) - f(a)}{h}, \quad (24)$$

concepto que aplicado al caso de (7) y (19), considerando el punto a evaluar $H_i$, resulta en

$$\tilde{Q}^{*\prime}(H_i) = \lim_{h \to 0}\frac{\tilde{Q}^*(H_i + h) - \tilde{Q}^*(H_i)}{h} \quad (25)$$

y

$$R'(H_i) = \lim_{h \to 0}\frac{R(H_i + h) - R(H_i)}{h}, \quad (26)$$

respectivamente, lo que permite reducir el número de puntos a ser evaluados.

Por otra parte, considerando el teorema del límite central para el estimador de Whittle (Fox and Taqqu, 1986; Dahlhaus, 1989), se tiene que en el caso de estimar un único parámetro $H$, si $\hat{H}$ es el valor que minimiza la función $Q(H)$ y $H_0$ su valor real, entonces

$$(H - H_0) \to N(0, \sigma_H), \quad (27)$$

siendo

$$\sigma_H^2 = \frac{4\pi}{N}\left[\int_{-\pi}^{\pi}\left(\frac{\partial \log f(\lambda, H)}{\partial H}\right)^2 d\lambda\right]_{H=H_0}^{-1} \quad (28)$$

la expresión que permite calcular la varianza del estimador.

Luego, el valor de (28) puede aproximarse mediante una suma de frecuencias de Fourier haciendo $H_0 = \hat{H}$ y, al igual que antes, el cálculo de la derivada puede aproximarse mediante un cociente de diferencias para un incremento suficientemente pequeño $h$, salvo que en este caso el valor elegido para $\theta_1$ es indiferente puesto que se produce su cancelación. De esta forma, haciendo

$$g(\lambda, H) = \log f(\lambda, H) \quad (29)$$

para simplificar la notación, reemplazando en (28) se obtiene

$$\sigma_H^2 = \frac{4\pi}{N} \left[ \int_{-\pi}^{\pi} \left( \frac{\partial g(\lambda, H)}{\partial H} \right)^2 d\lambda \right]_{H=H_0}^{-1}, \quad (30)$$

y puesto que $H_0 = \hat{H}$, entonces

$$\sigma_H^2 = \frac{4\pi}{N} \left[ \int_{-\pi}^{\pi} \left( \frac{\partial g(\lambda, H)}{\partial H} \right)^2 d\lambda \right]_{H=\hat{H}}^{-1}, \quad (31)$$

es decir,

$$\sigma_H^2 = \frac{4\pi}{N} \left[ \sum_{k=1}^{N^*} \left( \frac{\partial g(\lambda_k, H)}{\partial H} \right)^2 \right]_{H=\hat{H}}^{-1}. \quad (32)$$

Luego, de (24) aplicada sobre (29) se tiene que

$$\left. \frac{\partial g(\lambda_k, H)}{\partial H} \right|_{H=\hat{H}} = \frac{g(\lambda_k, \hat{H}+h) - g(\lambda_k, \hat{H})}{h}, \quad (33)$$

es decir,

$$\left. \frac{\partial \log f(\lambda_k, H)}{\partial H} \right|_{H=\hat{H}} = \frac{\log f(\lambda_k, \hat{H}+h) - \log f(\lambda_k, \hat{H})}{h}, \quad (34)$$

expresión que al ser reemplazada en (32), resulta finalmente en la varianza buscada

$$\sigma_H^2 = \left[ \sum_{k=1}^{N^*} \left( \frac{\log f(\lambda_k, H+h) - \log f(\lambda_k, H)}{h} \right)^2 \right]^{-1}. \quad (35)$$

Se destaca que para el caso del método de Whittle local resulta todo mucho más sencillo puesto que simplemente se considera

$$\sigma_H^2 = \frac{1}{4m}, \quad (36)$$

tal y como lo establece (23).

## 5. Conclusiones

Los modelos de tráfico convencionales basados en procesos de Poisson, o en términos más generales, procesos con dependencia de corto alcance (SRD), no son capaces de describir con detalle el comportamiento de los flujos de tráfico en las actuales redes de computadoras de alta velocidad, en particular para el caso de las redes Ethernet conmutadas según el estándar IEEE 802.3-2005. Consecuentemente, se hace necesario replantear el estudio de los sistemas de carga considerando tráficos de entrada autosimilares, producto que la demanda del tráfico autosimilar impone nuevos requerimientos a la ingeniería de redes, en especial en lo referido a estrategias de buffering para los equipos activos, estimación del rendimiento y diseño de topologías.

Aunque los estimadores de máxima verosimilitud de Whittle son atractivos porque reemplazan la evaluación de la inversa de la matriz de varianzas y covarianzas del vector de observaciones por transformadas de Fourier del mismo vector, son costosos desde el punto de vista computacional, puesto que las funciones objetivo deben evaluarse en cada iteración por métodos numéricos. En este sentido se estima que este inconveniente puede, en parte, salvarse introduciendo un algoritmo que disminuya el número de puntos a evaluar. Es más, esta introducción no tan solo significa reducir el costo computacional como tal, sino que permite disponer de una nueva alternativa para ser considerada en el estudio que sobre las prestaciones de los actuales sistemas de comunicaciones, posee el tráfico con naturaleza fractal.

Se propone que un método de búsqueda por bisección aplicado sobre la derivada de la función objetivo en el dominio de interés [0.5, 1[ aprovechando su convexidad y por ello la presencia de un solo mínimo, permite disminuir la cantidad de puntos a evaluar y con ello mantener una relación aceptable de compromiso entre el sesgo y la calidad de las estimaciones.

**English Summary**

**Estimation of Hurst exponent in self-similar traffic flows.**

**Abstract**

In this paper it presents, develops and discusses the existence of a process with long scope memory structure, representing of the independence between the degree of randomness of the traffic generated by the sources and flow pattern exhibited by the network. The process existence is presented in term of a new algorithmic that is a variant of the maximum likelihood estimator (MLE) of Whittle, for the calculation of the Hurst exponent (*H*) of self-similar stationary second order time series of the flows of the individual sources and their aggregation. Also, it is discussed the additional problems introduced by the phenomenon of the locality of the Hurst exponent, that appears when the traffic flows consist of diverse elements with different Hurst exponents. The instance is exposed with the intention of being considered as a new and alternative approach for modeling and simulating traffic in existing computer networks.

*Keywords:*

Self-similarity, Hurst exponent (*H*), Maximum Likelihood Estimator (MLE), Whittle estimator, Long-scope memory, Long-range dependence, Telematic traffic modeling.